\documentstyle[aps,epsf,prb,preprint,tighten]{revtex}
\def\myfigure#1#2{{\leftskip=0.10753\textwidth \rightskip\leftskip\small
\begin{figure}\baselineskip=14pt plus 2pt minus 1pt
\centerline{#1}\nobreak\smallskip\nobreak #2\end{figure}}}
\global\firstfigfalse
\begin{document}
\draft
\title{Peak effect and its evolution with defect structure in
YBa$_2$Cu$_3$O$_
{7-\delta}$ thin films at microwave frequencies.}
\author{Tamalika Banerjee$^1$ \thanks{e-mail:tamalika@tifr.res.in}\and
D. Kanjilal$^2$ 
and R.Pinto$^1$ \thanks{e-mail:rpinto@tifr.res.in}}
\address{$^1$Department of Condensed Matter Physics and Materials Science,\\
Tata Institute of Fundamental Research, Homi Bhabha Road, 
Mumbai 400 005, India \\
$^2$Nuclear Science Centre, Aruna Asaf Ali Marg, New Delhi 110 067, India}
\maketitle
\begin{abstract}

The vortex dynamics in YBa$_2$Cu$_3$O$_{7-\delta}$
thin films have been studied at microwave frequencies. 
A pronounced peak in the surface resistance, R$_s$, is observed in
these 
films at frequencies of 4.88 and 9.55 GHz for magnetic fields varying from 0.2
T to 0.8 T. The peak is associated with an order-disorder transformation of the 
flux line lattice as the temperature or field is increased. The occurrence of the 
peak in R$_s$  is crucially dependent on the depinning frequency, $\omega_p$ and 
on the nature and
concentration of growth defects present in these films. Introduction of
artificial defects 
by swift heavy ion irradiation with 200 MeV Ag ion at a fluence of $4
\times 10^{10}$ ions
/cm$^2$ enhances $\omega_p$ and  suppresses the peak at 4.88 GHz but the peak at
 9.55 GHz remains unaffected. A second peak at lower
temperature has 
also been  observed at 9.55 GHz. This is related to twin boundaries from 
angular dependence studies of R$_s$. Based on the temperature variation of
R$_s$, vortex phase diagrams have been constructed at 9.55 GHz. 

\end{abstract}

The dynamics of vortices forming the flux line lattice (FLL) in the mixed state 
of type II superconductors has been a subject of great interest over the last fe
w years [1-3]. The competition between intervortex interactions and pinning by 
disorder or defects [4-6] gives rise to the widely studied peak effect (PE) 
phenomenon which is the occurrence of a peak in the critical current density,
J$_c$, below its superconducting-normal phase boundary. 
The earliest understanding of this
 phenomenon, involves the rapid softening of the elastic moduli of the FLL as 
H$_{c_2}$(T) is approached which allows pinning sites to distort the lattice
more strongly, leading to a sharp rise in the critical current density 
[7-9]. This is described in the Larkin-Ovchinnikov collective pinning model as a
 reduction in the correlation volume V$_c$ which is the characteristic size
over which the FLL is ordered and where the effective pinning force on the FLL 
is given by 
\begin{equation}\label{eq1}
BJ_c (H) = (n_p \langle f^2\rangle /V_c)^{1/2}
\end{equation}
where n$_p$ is the density of pinning sites, f is the elementary pinning
force parameter, 
$B$ is the magnetic induction and V$_c$ is the volume of Larkin
domain. This 
scenario is generally true for weak collective pinning (point defects and uncorre
lated defects) where n$_p$ $>>$ n$_v$ (n$_v$ is the vortex density). On the
other hand, there
 is increasing evidence that for strong dilute pinning (eg., twin planes in 
twinned crystals) where n$_p$ $<$ n$_v$ , the transition to disorder above the
peak temperature T$_p$, is more likely to be accompanied by a plastic motion
of the vortex lattice.
 
	Till date, the statics and dynamics of the FLL have been probed at low 
frequencies by measurements of the dc critical current density J$_c$  and
very few studies 
have been done in the radiowave or microwave regime. At  high frequencies
 (low currents) the vortices undergo reversible oscillations and are less 
sensitive to flux creep [10]. In this sense, the small probe current leads to 
enhanced
 dissipation, which would otherwise have occurred for either high dc transport 
current or a large external magnetic field. The directly accessible quantity 
characterizing this response is the surface impedance given as
$Z_s = R_s + iX_s $
where the surface resistance R$_s$ determines dissipation and yields
information about the vortex dynamics and X$_s$,  the surface reactance,  is
a measure of the London penetration depth.

  	The earliest theoretical model of the dynamics of the Lorentz force of 
alternating screening currents at these frequencies was given by Gittleman and 
Rosenblum [11] and thereafter refined considerably [12]. For vortices oscillating 
close to the minimum of the pinning potential and experiencing a restoring force
, $\kappa_p$, (determined by the curvature of the pinning potential),  the
equation of motion for a massless flux line, 
neglecting Hall and stochastic thermal force,  is given as  
\begin{equation}\label{eq2}
\eta \dot{x} + \kappa_p x = J \times \phi_0
\end{equation}
where $J(t)$ is the microwave driving current, $x$ is the displacement from
equilibrium and $\eta=\phi_0 H_{c_2} /\rho_n$  is the Bardeen-Stephen
viscous drag coefficient. The vortex impedance is thus given by  
\begin{equation}\label{eq3}
\rho_v=\frac{\phi_0 B}{\eta}\frac{1}{(1+i\omega_p/\omega)}
=\frac{\omega^2-i\omega\omega_p}{\omega^2 + \omega_p^2}
\frac{H}{H_{c_2}}\rho_n
\end{equation}
where $\omega_p$ is the depinning frequency. The characteristic depinning
frequency, $\omega_p =
\kappa_p/\eta $, separates the low frequency regime ($\omega <
\omega_p$) dominated by pinning with inductive response, 
from a high frequency regime ($\omega > \omega_p$) of free vortex flow with
dissipation.  In this limit [13] R$_s$ is given as
\begin{equation}\label{eq4}
R_s=\left ( \frac{2H}{H_{c_2}} \right )^{1/2}\left
(\frac{\omega\omega_p}{\omega^2
+ \omega_p^2}\right )^{1/2}
\end{equation}
 From the above expression we see that for $\omega < \omega_p$,  R$_s$
increases with $\omega$ whereas for
 $\omega >> \omega_p$ ›› the response reduces to that of a normal metal
with $R_s = \rho_n H/H_{c_2}$.

	Although there have been few studies of the vortex dynamics at microwave
 and radio frequencies there are no reports on the observation of PE at microwave 
or radio frequency in either low T$_c$ or high T$_c$ superconductors. However,
recently we have reported [14] the first observation of PE in
DyBa$_2$Cu$_3$O$_{7-\delta}$ (DBCO) thin 
films at a frequency of 9.55 GHz. The observation of the peak was explained 
within the Larkin Ovchinnikov scenario in the weak collective pinning regime. 
Furthermore, it was shown that $\kappa_p$ (or $\omega_p$), has a similar
temperature and field variation 
as J$_c$ and shows a peak near T$_c$ (H$_{c_2}$) which in turn reflects
as a minimum in 
R$_s$. We had also conjectured that the shift in the position of the peak in
R$_s$ with 
frequency arises due to the pinning of vortices (or bundles of vortices) in 
potentials of varying strength $\kappa_p$ causing a distribution in time
scales of the FLL.

	Pinning of the FLL due to material disorder (growth defects as 
dislocations, stacking faults, point and surface defects) plays an important 
role in the transport properties of type II superconductors.  Ion irradiation 
is a well-established method to increase the lattice defect concentration in a 
controlled way with homogenous spatial distribution over the sample area [15]. 
Whereas point defects are not so effective pinning centres because of their 
extremely small coherence length, cylindrical defects of radius equal to the 
coherence length act as strong pinning centres. The
 strong pinning provided by such columnar defects (CDs) completely alters the eq
uilibrium properties of a clean vortex state [16]. In this work, we show that
the peak in R$_s$ of YBa$_2$Cu$_3$O$_{7-\delta}$ (YBCO) thin films at 4.88 GHz 
caused by random uncorrelated defects is suppressed with the introduction
 of artificial correlated columnar defects by 200 MeV Ag ion irradiation while 
the peak at 9.55 GHz remains unaffected.  

For the present study, several c-axis oriented YBCO (T$_c$ = 92 $\pm$ 0.2
K) thin films
 (thickness 2500 \AA), were grown by pulsed laser deposition
technique on single 
crystal $\langle 100 \rangle $ LaAlO$_3$ substrates. For microwave
transmission studies, the films were 
patterned into linear microstrip resonators of width 175 $\mu$m and length
9 mm using UV 
photolithographic techniques.  Details of the microwave measurements and
 determination of R$_s$ have been described earlier [17]. DC magnetic field 
varying
 from 0.2 T up to 0.9 T was applied perpendicular to the film plane (parallel to
 the c-axis of the film) using a conventional electromagnet. Since the penetrati
on depth, $\lambda$, of YBCO thin films is in the range 1500-2500
\AA $\;$ for the temperature 
range 0-77 K [17], a major fraction of the YBCO microstrip resonator is driven 
into the mixed state at the microwave frequencies. The temperature instability 
during measurements was always $<$ 30 mK. Irradiation was carried out using the 15
U D Pelletron accelerator at Nuclear Science Centre, New Delhi using 200 MeV
$^{109}Ag
^{+14}$ ions at a fluence of $4\times 10^{10}$ ions/cm$^2$ (corresponding
to a matching field B$_{\phi} \sim $
 0.8 T). For this choice of ion species and energy (electronic energy loss =2.3 
keV/\AA) columnar defects (CDs) are created [18]. 

The variation of R$_s$ with temperature, before and after irradiation, 
at various magnetic fields at 4.88 GHz (corresponding to the fundamental 
excitation of the microstrip) is shown in Fig.1. At field values $>$ 0 T, R$_s$ 
is found to exhibit a peak followed by a sudden dip just before T$_c$. 
Irradiation with 200 M
eV Ag ions at a fluence of $4\times 10^{10}$ ions/cm$^2$ causes the peaks
to be suppressed. Fig.2 shows the temperature variation of R$_s$ 
at 9.55 GHz 
(first harmonic excitation of the microstrip). Here we observe a 
second peak at lower temperatures for fields 
$>$ 0 T. However, there is no suppression of peaks with 200 MeV Ag
irradiation.  A manifestation of irradiation induced disorder is reflected in an
increase in R$_s$ at field values upto the matching field of 0.8 T.  At the
matching field of 0.8 T, 
the effect of pinning by CDs far surpasses the effect of disorder 
introduced by irradiation and a drop in the R$_s$ values is observed.
 
In the weak collective pinning scenario, the motion of the vortices 
inside V$_c$ is small as compared to the overall 
pinning potential arising collectively from such pinning sites inside the 
Larkin domain. At this frequency, the restoring force $\kappa_p$ acting on
the vortices causes it to oscillate in 
the potential well which causes depinning of the vortices near T$_c$ (at
T$_{p1}$). This order-disorder transformation at T$_{p1}$ exhibits as a
single peak in R$_s$. However, at a higher frequency of 9.55 GHz, an
additional peak at lower temperature (T$_{p2}$) is observed along with the
peak at T$_{p1}$. It is believed
 that the origin of these multiple peaks could be various defect structures comm
only found in thin films. Thin films grown by laser ablation have various types 
of defects as uncorrelated defects like point defects, oxygen 
deficiencies or their clusters, secondary phase precipitates or correlated
defects like twin boundaries. The 
differences in the pinning interaction of these defect sites with the vortices
 causes the system to have a distribution in time scales and leads to various 
peak structures at different temperatures [4] when the measurement frequency is 
changed from 4.88 to 9.55 GHz. The observation of secondary peaks in low frequency
 J$_c$-T plots and their possible origin has been widely studied in
literature [19]. 
It is likely that the secondary peak (at T$_{p2}$) observed at 9.55 GHz
in our case
could be due to twin boundaries - a common defect structure in these thin films. 
An angular correlation of this peak at T$_{p2}$ with the magnetic field
was done at
 9.55 GHz and is shown in the top inset of Fig.2. It is observed that the peak at 
T$_{p2}$, is significantly suppressed when the external magnetic field
is perpendicular 
to the c-axis of the film. However, the peak at T$_{p1}$  does not
have any angular dependence with the 
external magnetic field. This indicates that the peak at
 T$_{p2}$ arises due to pinning by twin boundaries.  Recent
magneto-optical imaging and 
magnetization measurements have also pointed out that twin boundaries are easy 
paths for flux pinning [20,21].
 
The bottom inset in Fig.2 is a plot of 1/R$_s$ (normalized) vs. T at a
field value
 of 0.4 T. To understand the significance of this plot let us go back to the 
correlation of J$_c$ with $\omega_p$ and hence R$_s$ [14]. It is seen that the
variation of 1/R$_s$ (normalized) is similar to that of 
J$_c$ and shows a peak before T$_c$ (H$_{c_2}$). Ion irradiati
on and creation of correlated defect centres as CDs are well known to strongly p
in the flux lines in such defect sites. Liberating a vortex from CDs requires co
nsiderable amount of energy causing an increase in J$_c$. This is reflected in 
an increase in 1/R$_s$ (normalized)  in our case. 

Within the collective pinning scenario for weak pinning centres, the model 
variation of depinning frequency $\omega_p$ with T/T$_c$, which shows a minimum
followed by a peak at T$_p$,
 critically depends on the measurement
frequency ( as proposed in our earlier paper [14]), curve as shown by `a'
in Fig.3. 
Thus, at a frequency 4.88 GHz (where
$\omega < \omega_p^{dip}$ or $\omega_1 $ in curve `a' of Fig.3), the peak
observed is less pronounced as compared to that at 9.55 GHz 
(where $\omega_p^{peak} > \omega_2 > \omega_p^{dip}$, curve `a' in Fig.3).
Artificially introduced 
defects like columnar defects lying in the vicinity of the other scattered 
defects enhance their pinning strength and prevent a flux line to be depinned 
at 4.88 GHz from such sites. In other words, $\omega_p$ of the YBCO film is 
increased above 4.88 GHz which is depicted in the upward shift of the model plot
 (curve `b' of 
Fig.3). However, the irradiation induced enhancement in $\omega_p$ is insufficient 
to keep the vortices pinned at 9.55 GHz; hence there is no suppression of $T_
{p1}$ after irradiation. 

Based on the temperature variation of R$_s$ (at 9.55 GHz), tentative vortex phase 
diagrams have been constructed and shown in Figs.
4a \&b for the pristine and irradiated sample, respectively. The onset of
order-disorder transformation of the FLL indicated by T$_{onset}$ ( onset of peak
in R$_s$)  corresponds to a minimum in $\omega_p(\omega_p^{dip})$ . The
process of disordering is complete at 
the peak in $\omega_p(\omega_p^{peak})$ which corresponds to the minimum
of R$_s$ at T$_{p_1}$. Hence, the 
phase diagram shown comprises an ordered vortex state, which crosses over to 
a fully disordered state 
via a partially ordered vortex state as the
temperature or field is increased. The crossover region, or the partially
ordered vortex solid region, broadens out over a greater
temperature range in the case of the irradiated sample. 
The subsequent transformation to a totally disordered or normal phase
is indicated by
T$_c$ (deduced from microwave data, R$_s \sim 10 m \Omega$).
This indicates that irradiation introduces disorder in the system which
modifies the dynamics of the vortex lattice in a major way.

In summary, we have studied peak effect at microwave frequencies 
(4.88 and 9.55 GHz) in YBCO thin films for H // c before and after irradiation
with 200 MeV Ag ions at a fluence of $4\times 10^{10}$ ions/cm$^2$ . 
We have shown that the peak in R$_s$ at such
frequencies critically 
depends on the nature and concentration of defects. A suppression of the peak 
at 4.88 GHz is observed after irradiation due to enhanced pinning by the 
columnar defects which 
shifts the $\omega_p$ vs. T/T$_c$ model plot (proposed in our earlier work) 
to higher $\omega_p$ values. Angular-dependence measurements of R$_s$ with 
magnetic field indicate
that the second peak at T$_{p2}$ observed at 9.55 GHz, is related to the twin
boundaries. From the phase diagrams, based on the temperature variation of R$_s$
 at 9.55 GHz,  we see that irradiation induced disorder broadens
 the crossover region between the ordered and the disordered phase. We have 
carried out our analysis using the Gittlemann-Rosenblaum model 
where the effects of thermally activated flux motion are neglected. 
It would be interesting to study the PE phenomenon and the vortex dynamics 
by using a more refined model where the
 role of thermal effects, at these frequencies, are incorporated. 

It is a pleasure to acknowledge useful discussions with Shobo Bhattacharya
and Pratap Raychaudhuri. Authors wish to thank S.P. Pai and Avinash
Bhangale for experimental assistance.

\newpage
\myfigure{\epsfysize 3.3in\epsfbox{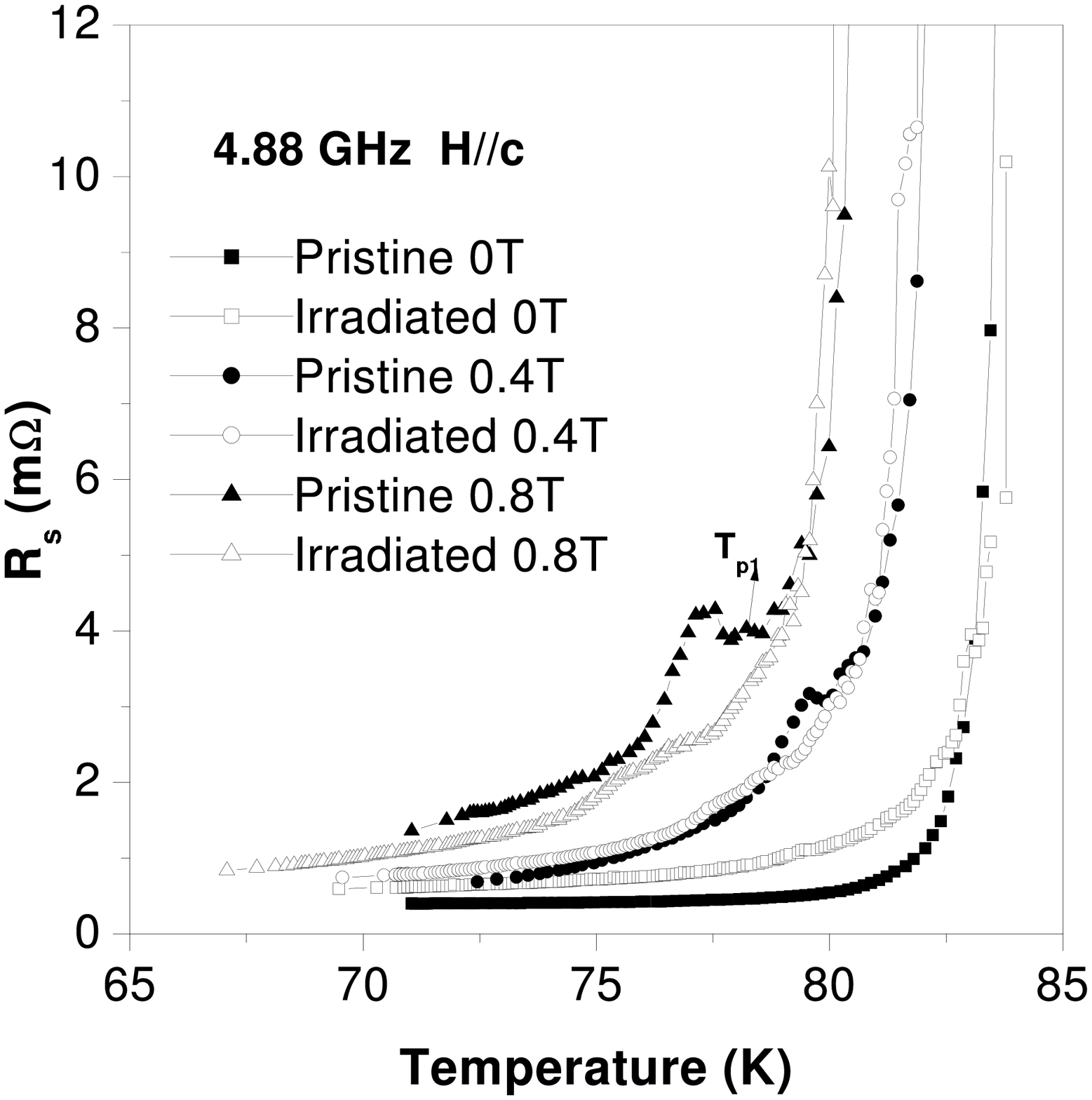}}{
Fig.1 : R$_s$ vs. T plots at 4.88 GHz for various applied fields (// c) for both
 pristine and irradiated films.}

\myfigure{\epsfysize 3.3in\epsfbox{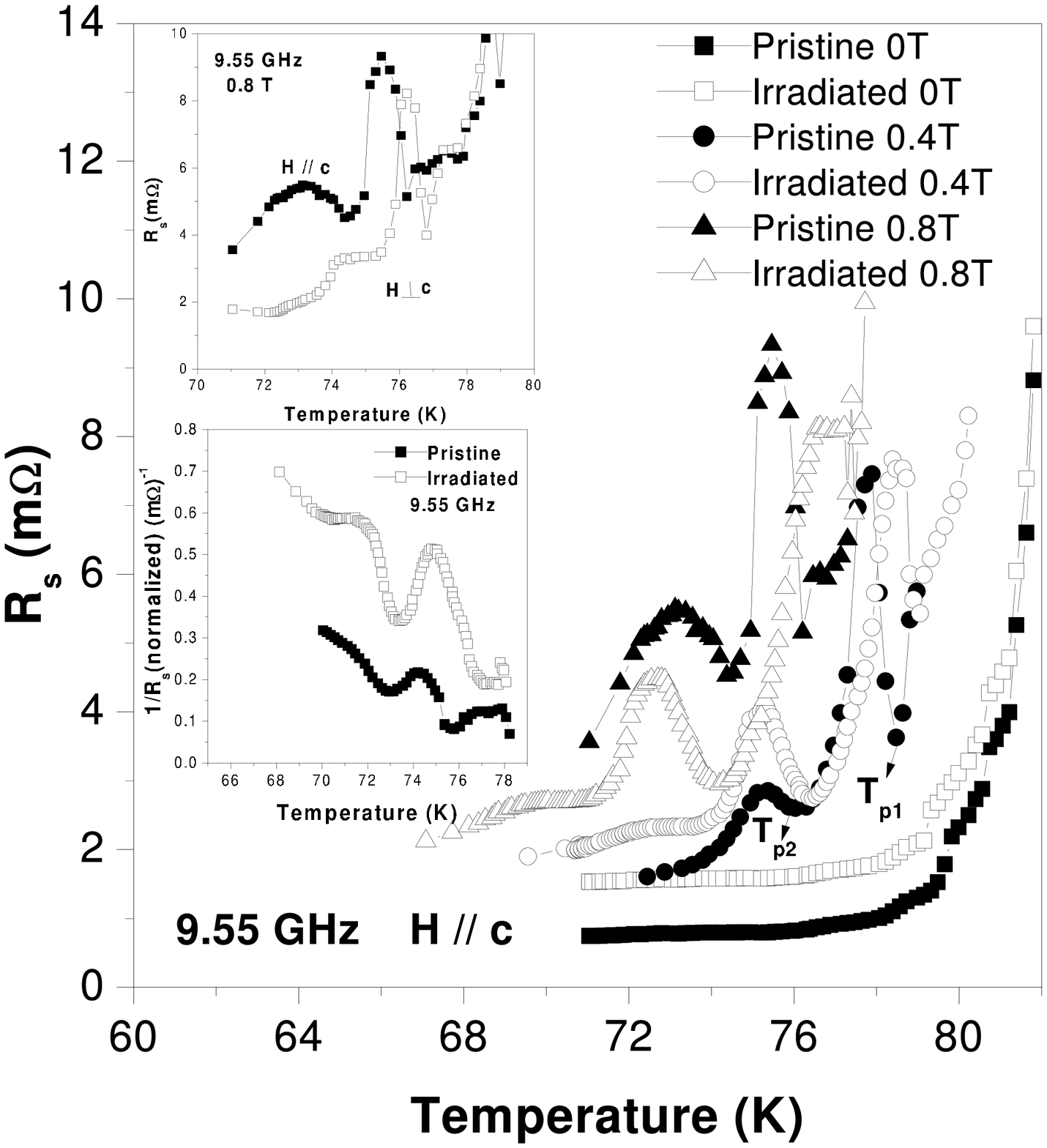}}{
Fig.2: R$_s$ vs. T plots at 9.55 GHz for various applied fields (// c) for both 
pristine and irradiated films. Top Inset shows R$_s$ vs. T plot at 0.8T for H // c 
and  H $\perp$ c at 9.55 GHz.
Bottom Inset shows 1/R$_s$ (normalized) vs. T plots at 0.4 T for H // c at
9.55 GHz.}

\myfigure{\epsfysize 2.5in\epsfbox{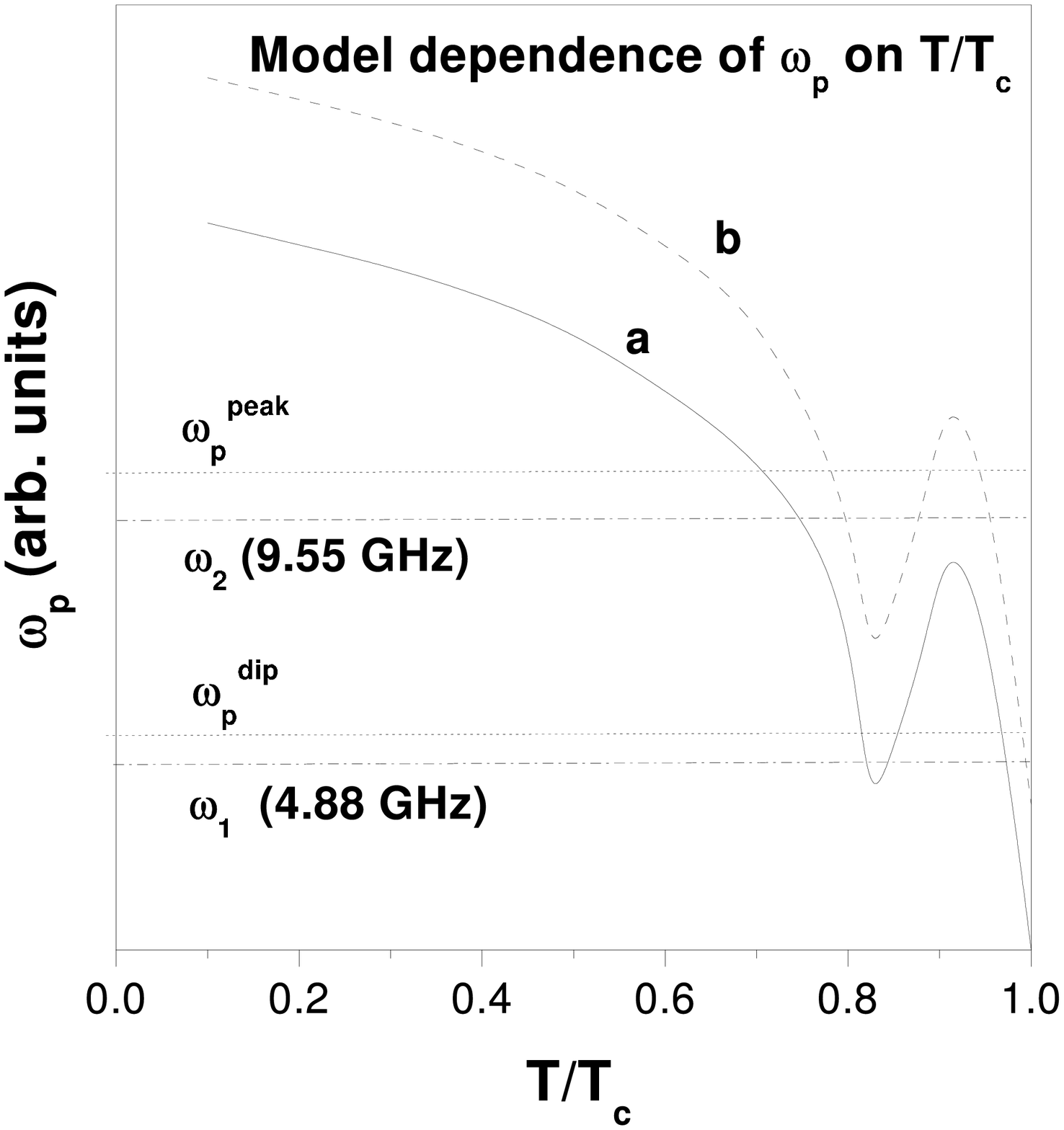}}{
Fig.3 Model dependence of $\omega_p$ on T/T$_c$ for (a) pristine sample
and (b) the
sample irradiated with 200 MeV Ag ions at a fluence of
$4 \times 10^{10}$ ions/cm$^2$.}

\myfigure{\epsfysize 2.3in\epsfbox{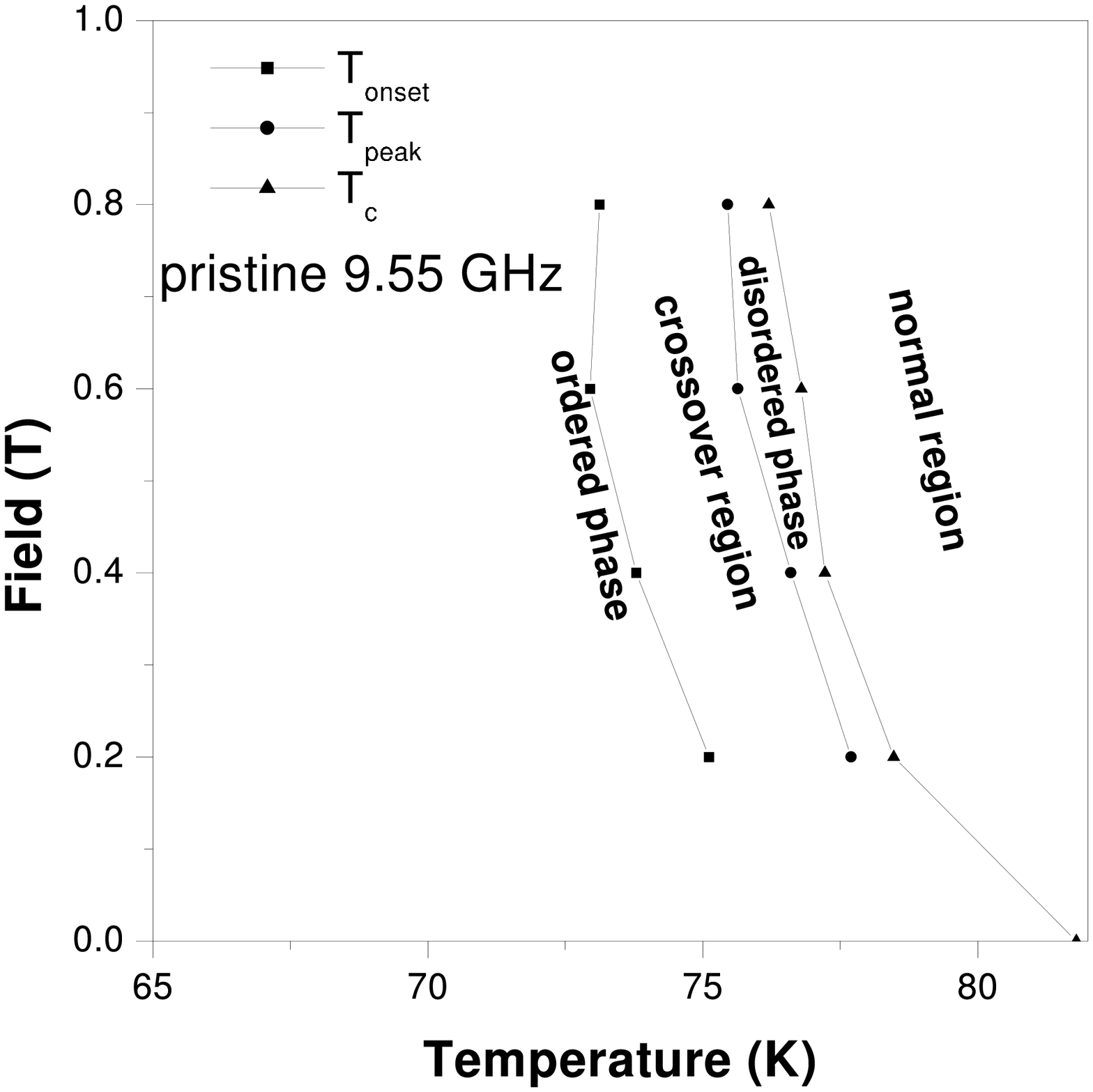}}{}
\myfigure{\epsfysize 2.3in\epsfbox{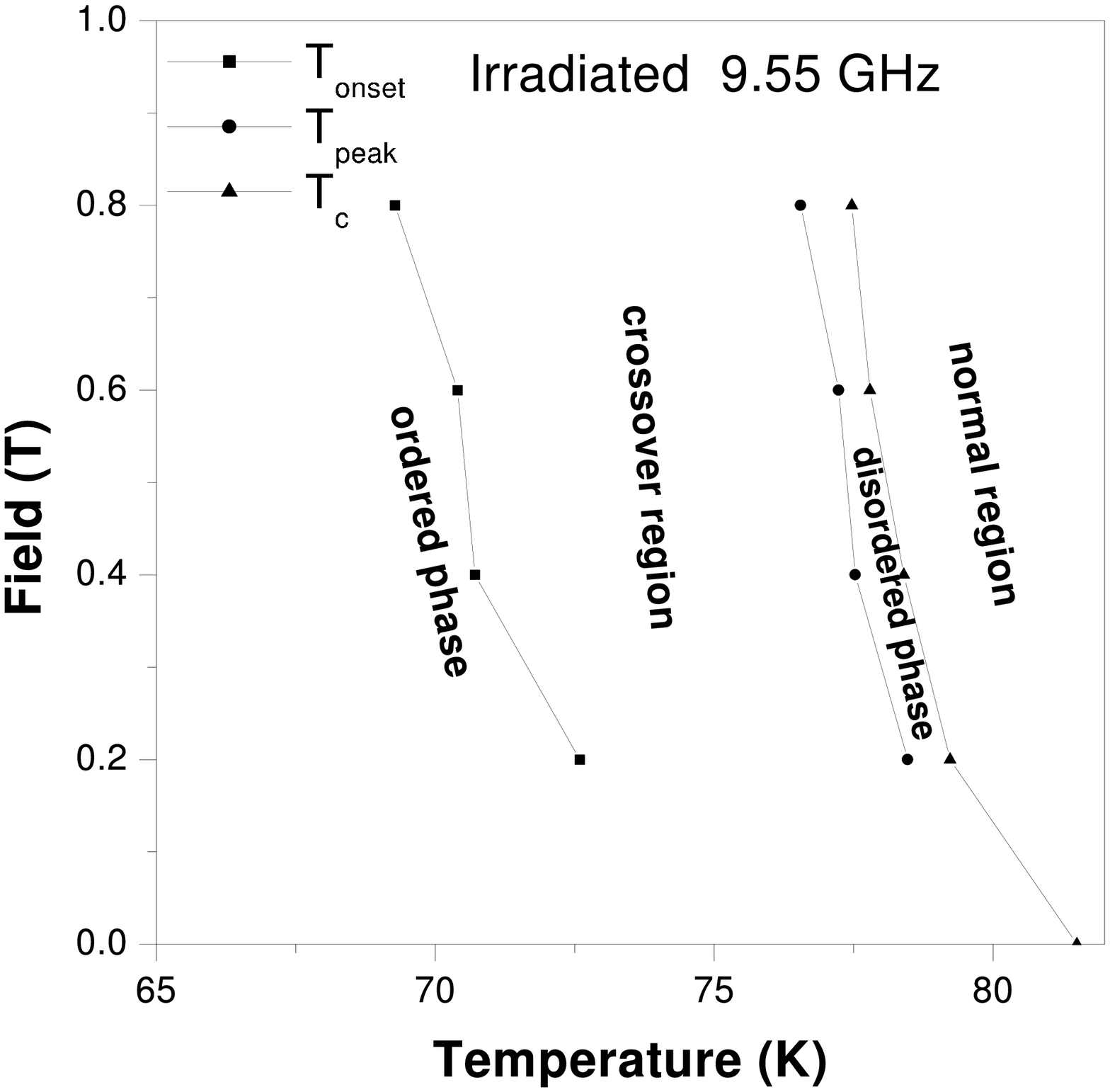}}{
Fig.4 . Phase diagrams:
(a) At 9.55 GHz, Pristine sample
(b) At 9.55 GHz, Irradiated  sample.}

\end{document}